# Straight Directional Couplers via Scan-Engineered Index Control

MOHAN WANG,[1] MARTIN J. BOOTH,[1] PATRICK S. SALTER[1, *]

[1]Departmnent of Engineering Science, Parks Road, University of Oxford, Oxford, UK OX1 3PJ
*patrick.salter@eng.ox.ac.uk

**A novel design for straight directional waveguide couplers and interferometers is demonstrated in glass, fabricated using femtosecond laser direct writing and operating at telecommunication wavelengths (~1550 nm). The devices consisted of parallel waveguides with a spacing of 15 μm, where the coupling strength was controlled by scan-engineered refractive index modulation along the length of the waveguide. Using this approach, we realized a 50:50 directional coupler formed by two identical waveguides with a footprint of < 40 μm × 15 μm × 6 mm, as well as a Mach–Zehnder interferometer with unbalanced arms. A waveguide array with 15 μm spacing was also demonstrated, highlighting the potential for compact, high-density, and three-dimensional photonic integration.**

Integrated photonics have emerged as a key platform for applications including quantum information processing [1], artificial intelligence hardware [2], biosensors [3], and optical interconnects in data centers [4]. They enable miniaturized devices with relaxed alignment requirements, reduced system complexity, and improved operational stability. Femtosecond laser direct writing is one of the enabling fabrication technologies. By exploiting the nonlinear interactions between a tightly focused laser beam and a transparent dielectric, it enables the fabrication of truly three-dimensional (3D) photonic circuits without the need for cleanroom facilities. This approach supports rapid prototyping while producing low-loss waveguides (< 0.1 dB/cm) with mode profiles well matched to optical fibers [5,6].

Directional couplers (DCs) are fundamental building blocks in integrated photonics, enabling power splitting and combining, and forming the basis of interferometric circuits such as Mach–Zehnder interferometers (MZIs). The coupling between two waveguides is governed by coupled mode theory, where the coupling strength at a certain wavelength depends on the overlap of evanescent fields and the propagation constant difference. In conventional DCs (Fig. 1), coupling is mostly controlled by the spatial separation between waveguides. The device typically consists of three regions: a transmission region with negligible coupling, a coupling region where the waveguides are brought into close proximity to enable power exchange, and a transition region that connects the two, using S-bend geometries to ensure low-loss mode evolution.

Laser-written waveguides typically exhibit a low refractive index contrast ($\sim 10^{-3}$), resulting in weak optical confinement [6–8]. As a result, the port-to-port separation in laser-written directional couplers is typically on the order of 100–200 μm to suppress unwanted evanescent coupling, while the overall device length is dominated by the transition regions. This is due to the requirement of large bending radii to minimize propagation loss. A bend loss of $\sim$ 0.3 dB/cm requires a bending radius of $\sim$ 30 mm, leading to device lengths of 8–24 mm [9–11], limiting the device size and integration density.

One approach to reduce device size is to increase the refractive index contrast, thereby improving optical confinement and enabling tighter bends. For instance, Ross-Adams *et al.* demonstrated a two-step fabrication process, combining high repetition rate writing with subsequent low repetition rate multi-scan writing, achieving a refractive index contrast of 0.0112 and a bend loss of 1.0 dB/cm at a radius of 3.5 mm [5]. Similarly, adaptive optics-assisted laser writing has enabled multi-scan waveguides with refractive index contrasts up to ~0.017, achieving low bend loss for smoothly curved waveguides with radii on the order of tens of millimeters [12,13]. While these approaches improve bending performance, the device footprints remain constrained by the need for curved transition regions.

Alternatively, straight directional couplers have recently been proposed to eliminate bending entirely. In these designs, the waveguide separation remains constant, and coupling is controlled via insertion of extra laser write tracks in the coupling region to promote mode mismatch along the propagation direction. Yan *et al.* demonstrated such devices using asymmetric waveguides with different propagation constants, achieving low crosstalk at a spacing of 9 μm [14]. Similar concepts have been extended to multi-port devices such as tritters [15]. Nevertheless, the approach requires dissimilar waveguide geometries and careful design of propagation constant mismatch, which increases device complexity and limits scalability.

In this work, we demonstrate a straight directional coupler and interferometer formed by identical waveguides with a spacing of 15 μm, in which coupling was controlled by scan-engineered refractive index modulation to generate adiabatic mode conversion along the propagation direction. This approach eliminated the need for both bending and geometric asymmetry, enabling compact and scalable photonic devices.

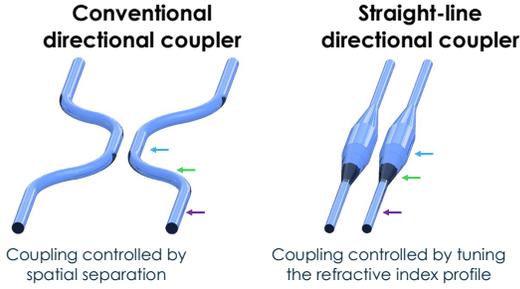

Fig. 1. Conventional DCs (left) and an index controlled straight DC (right), where coupling is controlled by scan-engineered refractive index profile. The three arrows point to the transmission region (purple), the transition region (green), and the coupling region (blue).

A femtosecond laser system (Pharos SP06-1000-PP) was used for fabrication, delivering a second harmonic generated wavelength of 515 nm with a pulse duration of 170 fs and a repetition rate of 1 MHz. The laser power was controlled using a half-wave plate and a polarizing beam splitter, for a pulse energy of 50 nJ. The output beam was expanded using a telescope and projected onto a spatial light modulator (Hamamatsu X10468) to compensate for the depth dependent spherical aberration [16]. The beam was then focused onto the pupil plane of a 20×, 0.5 NA objective via a second telescope set up.

The fused silica glass substrate was mounted on a three-axis motion stage (Aerotech *ABL10100L* and *ANT95-3-V*), and waveguides were inscribed at a speed of 4 mm/s. The laser was linearly polarized, with the polarization direction perpendicular to the writing direction. Waveguides were inscribed at a depth of 300 μm below the substrate surface, although similar performance is anticipated at other depths [17].

Multi-scan waveguides were fabricated using a 4 × 6 array of laser-written tracks, with the scan density continuously varied along the propagation direction. A schematic of the writing method is shown in Fig. 2(a), where each blue line represents a single laser-written track. The horizontal spacing ($\Delta X$) was varied from 0.75 μm in the transmission region to 2.25 μm in the coupling region, while the vertical spacing ($\Delta Y$) was varied from 0.5 μm to 1.5 μm. Each individual track has a typical dimension of approximately 1 × 3 μm². The scans exhibited significant overlap between adjacent tracks, forming a continuous waveguide structure.

A taper region was introduced between the transmission and coupling regions to enable adiabatic mode conversion [12]. Cross-sectional photos of the taper region are shown in Fig. 2(b), where the linearly decreasing scan density from the transmission region to the coupling region was reflected in the changing waveguide shape, leading to a variation of the effective refractive index along the propagation direction [18].

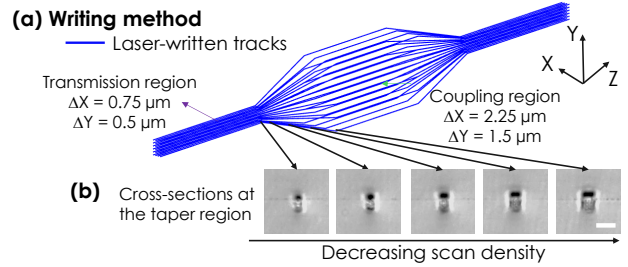

Fig. 2. (a) Schematic of the writing method for each waveguide in a straight DC. Each blue line indicates a laser-written track. The waveguide consisted of 24 tracks with varying scan density along the propagation direction and (b) photos of the waveguide cross sections in the taper region, showing the gradual transition in the scan density. Scale bar represents 15 μm.

The mode field of the waveguides at 1550 nm was characterized by butt-coupling light from a single-mode telecom fiber mounted on a six-axis nano-positioning stage and imaging the output using an InGaAs camera. Figure 3 shows the optical micrographs and corresponding mode profiles of the high-density and low-density waveguides.

The high-density waveguide exhibited a cross-sectional area of approximately 6 × 7 μm², while the waveguide in the coupling region had a larger cross-section of approximately 12 × 12 μm². The mode field diameters (MFDs) were extracted from the measured profiles and are indicated in the figure. The high-density waveguide measures MFDs of 8.1 and 7.9 μm in the horizontal and vertical directions, respectively. When interfaced with a single-mode telecom optical fiber, a mode mismatch loss of approximately 0.3 dB was calculated, which could be further reduced using an input taper [12]. The measured MFDs of the low-density waveguide are 12.0 μm and 10.7 μm in the horizontal and vertical directions.

The refractive index contrast was estimated from the near-field mode profile and the diffraction pattern to be 0.010 ± 0.002 and 0.002 ± 0.001 for the high-density multi-scan waveguide and the low-density multi-scan waveguide, respectively [19].

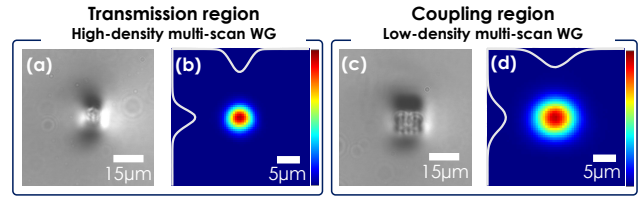

Fig. 3. (a–b) Cross-sectional photo (a) and 1550 nm mode field (b) of a high-density multi-scan waveguide, from the transmission region; (c–d) cross-sectional photo (c) and mode field (d) of the low-density waveguide, from the coupling region.

A waveguide-to-waveguide spacing of 15 μm was selected based on numerical simulations. Figure 4(a) shows the schematic of a horizontal straight directional coupler. Devices were fabricated with a taper length of $L_T = 1$ mm, while the coupling length $L_C$ was varied from 0.5 to 5.25 mm. The corresponding power splitting ratios were measured, covering a full range from 0 to 100%. A 50:50 splitting ratio was obtained at $L_C = 3.35$ mm, corresponding to a total device length of 5.35 mm. A microscope image of the output facet is shown in Fig. 4(b), and the measured mode profiles for excitation at each input port are presented in Fig. 4(c) and (d), yielding a splitting ratio of 47:53.

A control experiment was performed to verify negligible coupling between waveguides with different scan densities [Fig. 4(e)]. The measured mode profiles [Fig. 4(f,g)] indicate a crosstalk of approximately 0.05 dB between the high-density and low-density waveguides.

To demonstrate three-dimensional integration capability, a vertical directional coupler was also fabricated, as shown in Fig. 4(h–j). The taper and coupling lengths were determined using the same approach as for the horizontal device. A design with $L_T = 0.5$mm and $L_C = 4$mm achieved a 50:50 splitting ratio with a total device length of 5 mm.

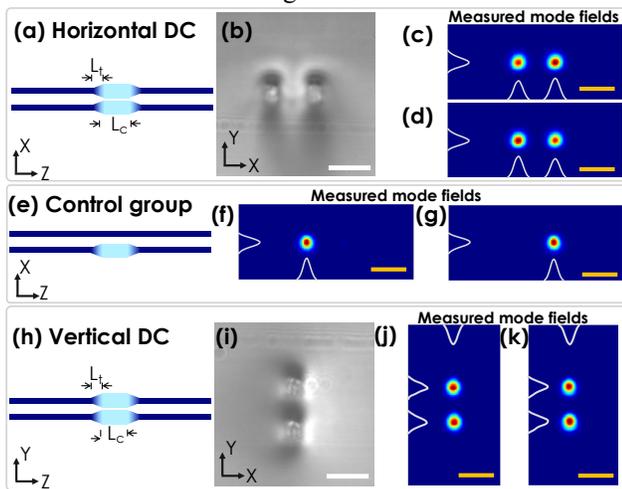

waveguides remained largely unaffected. A power splitting ratio of 52:48 was obtained.

A vertical directional coupler was also implemented using the same fabrication strategy. In this case, the coupler connected the waveguides at Row 2, Column 2 and Row 3, Column 2, as shown in Fig. 5(c). The measured output mode profiles (Fig. 5(d)) confirmed similar coupling behavior, with a splitting ratio of 49:51 and minimal perturbation to neighboring waveguides.

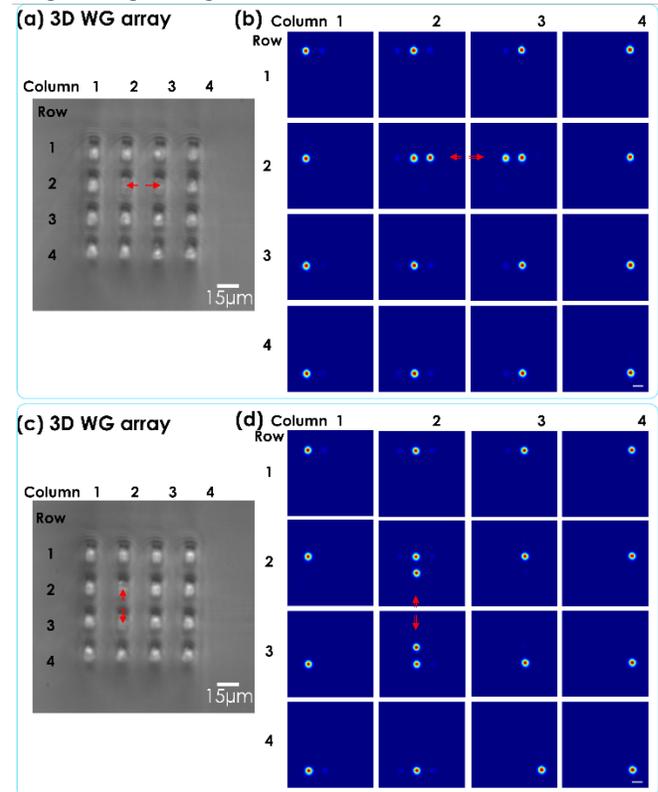

Fig. 4. (a) Design of a horizontal straight DC; (b) microscopic image of the end facet of a fabricated device; (c–d) measured 1550 nm mode fields when light was injected into the left (c) and right (d) input ports; (e) design of a vertical DC; (f) image of the end facet; (g–h) measured mode fields when light was injected into the top (g) and bottom (h) ports; (k–m) a comparative experiment in which only one waveguide included a coupling region, showing no coupling between the high-density and low-density waveguides; (l–m) measured mode fields when light was injected into each port. The white and orange scale bars represent 15 µm.

An array of 16 straight waveguides was fabricated in a 10 mm-long glass substrate, with a uniform center-to-center spacing of 15 µm in both horizontal and vertical directions. Figures 5(a) and (c) show the output facets of two such arrays. All waveguides were designed as high-density waveguides, except for a selected pair in the central region that were interconnected via DCs.

In the configuration shown in Fig. 5(a), a horizontal DC connected the waveguides at Row 2, Column 2 and Row 2, Column 3. The corresponding output mode profiles for excitation at each input port are shown in Fig. 5(b), demonstrating controlled power transfer between the coupled waveguides. Despite minor crosstalk, comparable to that observed in Fig. 1(f), light propagation in the surrounding

Fig. 5. The output facets (a) and (c) of a waveguide array with equal spacing of 15 µm in both directions written on a 1-cm substrate. In the design shown in (a), a DC was placed between the waveguides at Row 2, Column 2 and Row 2, Column 3 (indicated by red arrows), and in (c), between Row 2, Column 2 and Row 3, Column 2 (also indicated by red arrows). (b) and (d) show the measured mode fields when light was coupled into each port; the waveguides connected by DCs are indicated by dashed double-line arrows. The intensities have been normalized in each sub-image. Scale bars in the bottom right images: 15 µm.

One application of directional couplers is cascaded power splitting for multi-port optical distribution, such as coupling light into a multi-channel V-groove fiber array. The capability of the proposed straight-line directional coupler was further demonstrated using a cascaded configuration to achieve 1×4 power splitting. A schematic of the design is shown in Fig. 6(a). The device was fabricated with each DC feature the same design as the one shown in Fig. 4(a). Figure 6(b) shows the end facet photo. The measured output mode profiles for excitation at the input port indicated by the red arrow are shown in Fig. 6(c), demonstrating a near-uniform power distribution of 26:26:25:23 across the four output ports.

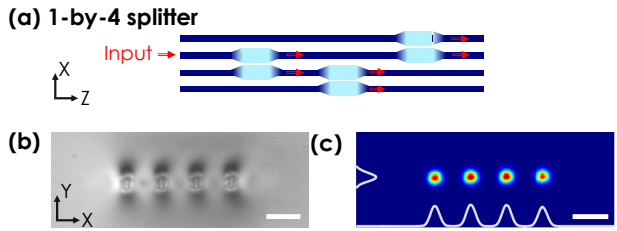

Fig. 6. (a) Schematic of a cascaded DC, (b) image of the device end facet, and (c) the mode field when light was injected into the port indicated by the red arrow in (a). Scale bars: 15 µm.

Mach–Zehnder interferometers (MZIs) convert phase differences between two optical paths into measurable intensity variations, enabling applications in modulation, switching, and sensing. A straight-line MZI can be realized by cascading two straight directional couplers, as illustrated in Fig. 7(a,b). The devices were fabricated on a 20 mm-long substrate. An unbalanced MZI (Fig. 7(b)) was implemented by introducing an optical path difference (OPD) between the two arms using high-density and low-density multi-scan waveguides. The designed OPD was 17.6 mm. The transmission spectrum, shown in Fig. 7(c), was measured using a tunable laser source and photodetector system. A free spectral range (FSR) of 15.5 nm was obtained, corresponding to an estimated refractive index difference of ~0.008, in good agreement with the previous refractive index estimation.

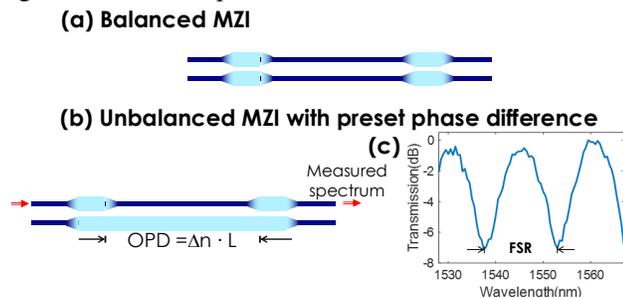

Fig. 7. (a) Schematic of a balanced MZI; (b) schematic of an unbalanced MZI, with the optical path difference defined by the arm length difference $L$; (c) experimentally measured transmission spectrum of a fabricated MZI following the design in (b), with a length difference of 17.6 mm between the two arms.

In this work, we have demonstrated a straight-line directional coupler and interferometric platform based on scan-engineered refractive index modulation in femtosecond laser-written waveguides. By controlling the refractive index through multi-scan fabrication, coupling between identical waveguides can be precisely tuned without relying on large bending radii or geometric asymmetry. This work provides a flexible and scalable method for engineering coupling and phase in femtosecond laser-written photonic circuits, enabling compact, high-density, and three-dimensional integrated photonic systems.

**Funding.** Advanced Research + Invention Agency (NACB-SE02-P01); Q1 Engineering and Physical Sciences Research Council (EP/W025256/1).

**Disclosures**. MB:OpsydiaLtd(I,C,S).PS:OpsydiaLtd(I,C,S).